\documentstyle[12pt]{article}

\begin{document}
\begin{titlepage}
\begin{center}

Oct  31, 2001     \hfill    LBNL-46942-Rev  \\

\vskip .5in

{\large \bf Bell's Theorem Without Hidden Variables}
\footnote{This work was supported in part by the Director, Office of Science, 
Office of High Energy and Nuclear Physics, of the U.S. Department of Energy 
under Contract DE-AC03-76SF00098.}

\vskip .50in
Henry P. Stapp\\
{\em Lawrence Berkeley National Laboratory\\
      University of California\\
    Berkeley, California 94720}
\end{center}

\vskip .5in

\begin{abstract}

Experiments motivated by Bell's theorem have led some physicists to 
conclude that quantum theory is nonlocal. However, the theoretical 
basis for such claims is usually taken to be Bell's Theorem, which shows 
only that if certain predictions of quantum theory are correct, and 
a strong hidden-variable assumption is valid, then a certain locality 
condition must fail. This locality condition expresses the idea that what 
an experimenter freely chooses to measure in one spacetime region can have 
no effect of any kind in a second region situated spacelike relative 
to the first. The experimental results conform closely to the predictions 
of quantum theory in such cases, but the most reasonable conclusion to 
draw is not that locality fails, but rather that the hidden-variable 
assumption is false. For this assumption conflicts with the quantum 
precept that unperformed experiments have no outcomes. The present 
paper deduces the failure of this locality condition directly from 
the precepts of quantum theory themselves, in a way that generates no
inconsistency or any conflict with the predictions of relativistic 
quantum field theory.

\end{abstract}
\medskip
\end{titlepage}

\newpage
\renewcommand{\thepage}{\arabic{page}}
\setcounter{page}{1}

{\bf 1. Introduction.}
\vspace{.1in}

A recent issue of  Physics Today[1] has a bulletin entitled 
``Nonlocality Get More Real''. It reports experiments at three 
laboratories (Geneva [2], Innsbruck[3], and Los Alamos[4]) directed 
at closing loopholes in proofs that quantum phenomena cannot be 
reconciled with classical ideas about the local character of the 
physical world. 

The experiment reported in the first of these papers [2] confirms
the existence of a classically unexplainable connection between phenomena
appearing at essentially the same time in two villages separated 
by a distance of more than 10km, and the paper begins with the provocative
statement ``Quantum theory is nonlocal.''  The longer version [5]  
says ``Today, most physicists are convinced that a future loophole-free 
test will definitely demonstrate that nature is indeed nonlocal.''

The theoretical basis for such claims is usually taken to be Bell's Theorem, 
which, however, shows only that if certain predictions of quantum theory are 
correct, {\it and if a certain hidden-variable assumption is valid,} then 
a locality condition must fail. This locality condition expresses the 
physical idea, suggested by the theory of relativity, that what an 
experimenter freely chooses to measure in one spacetime region can have 
no effect of any kind in a second region situated spacelike relative to 
the first. 

The experimental results conform closely to the predictions of quantum 
theory in such cases, but the most natural conclusion to draw is not 
that locality fails, but rather that the hidden-variable assumption is 
false. For the ``hidden-variable assumption'' of Bell's original 
theorem[6] is basically the assumption that a set of possible outcomes 
can be simultaneously defined for each of the alternative mutually 
incompatible experiments that the experimenters might choose to perform.
This assumption violates the precepts of quantum philosophy by 
assigning definite, albeit unknown, values to the outcomes of 
mutually incompatible measurements.

Bell[7] later introduced a seemingly weaker local hidden-variable
assumption, but it can be shown[8,9] that this later form entails the 
original one, apart from errors that tend to zero as the number of 
experiments tends to infinity. Thus both forms of the hidden-variable 
assumption contradict the basic quantum precept that one cannot, 
in general, consistently assign possible outcomes to unperformed 
measurements. Hence, from the viewpoint of orthodox quantum theory, 
Bell's hidden-variable assumption, in either form, is more likely to 
fail than the locality condition. Consequently, these hidden-variable 
theorems provides no adequate basis for the  claim that 
``Quantum theory is nonlocal'', or that ``nature is indeed nonlocal.'' 
 
The present paper describes a fundamentally different kind of proof.
It shows that a violation of the locality condition described above
follows logically from assumptions that formalize the precepts 
of orthodox of quantum theory itself. This proof is logically rigorous. 
But like all proofs and theorems of this general kind it is based 
essentially on the use of statements about contrary-to-fact situations. 

I have just mentioned that the basic principles of quantum philosophy
forbid the unrestricted use of contrary-to-fact, or counterfactual 
notions. So it might be thought that the new proof is basically no 
better than the ones employing hidden-variables. However, there is an
essential logical difference between the two proofs: Bell's 
hidden-variable assumption was an ad hoc assumption that had no 
foundation in the quantum precepts. Indeed, it {\it directly 
contradicted the quantum precepts.} Moreover, it led, 
when combined with the predictions of quantum theory, to {\it 
logical inconsistencies}. This latter fact vindicates the quantum 
precept that bans the hidden-variable assumption, and thus undermines
the significance of any conclusions derived from that assumption.  

The proof to be presented here differs from the hidden-variable 
theorems in two crucial ways. First, as just mentioned, the assumptions 
are {\it direct expressions} of the precepts of orthodox quantum precepts 
themselves, rather than being, as regards the hidden-variable aspect, a 
direct violation of those precepts. Second, the assumptions used in the 
present proof, lead only to a very restricted set of true counterfactual 
statements, and these lead neither to any logical contradiction, nor to 
any conflict with the predictions of relativistic quantum field theory. 
Thus the proof to be presented here lies on a logical level different 
from that of the proofs that follows Bell's hidden-variable approach.

What assumptions are used in the present approach?

The first assumption is that, for the purposes of understanding
and applying quantum theory, the choice of which experiment
is to be performed can be treated as a free variable. Bohr
repeatedly stressed this point, which is closely connected to
his ``complementarity'' idea that the quantum state contains
complementary kinds of information pertaining to the various
alternative mutually exclusive experiments that might be chosen.
Of course, only one {\it or} the other of two mutually incompatible
measurements can be actually performed, not both, and only a 
measurement that is actually performed can be assumed to 
have a definite outcome.

It is worth recalling in this connection that Bohr did not 
reject the argument of Einstein, Podolsky, and Rosen out of hand,
simply because it involves considering two mutually incompatible 
experiments. Bohr accepted that feature of the argument, and was 
therefore forced to find another, much more subtle, ground for 
rejecting the argument of those authors, which, like the one to be
presented here, but unlike those of Bell, scrupulously avoids any 
direct assumption that an unperformed experiment has a definite 
outcome.

The second assumption of the present work is that an outcome 
that has appeared to observers in one spacetime region, and has been
recorded there, can be considered to be fixed and settled by the time 
the observations and recordings are completed: this outcome is treated
as being independent of which experiment will be freely chosen and 
performed by another experimenter acting in a spacetime region {\it 
lying later than} the first region. This no-backward-in-time-influence 
property is assumed to hold in {\it at least} one Lorentz frame, 
which I shall call LF.

These two assumptions immediately entail the truth of a very limited
class of counterfactual statements. These are statements of the form: 
If experiments $E$ and $F$ are freely chosen and performed in earlier and 
later regions, respectively, and the outcome in the earlier region is $E+$, 
then the outcome in the earlier region would (still) be $E+$ if everything 
in nature were the same except for (1), {\it a different free choice} 
made in the later region, and (2), {\it the possible consequences} of 
making that alternate later free choice {\it instead of} the actual free 
choice. The no-backward-in-time-influence condition asserts that the 
possible consequences of the experimenter's making the 
alternate (i.e., counterfactual) later free choice do not include any 
change of an outcome that has already appeared to observers in the 
earlier spacetime region, and has been recorded there. 

This assumption of no backward-in-time action is the expression of 
a theoretical idea: it can never be empirically confirmed. 
On the other hand, it is completely compatible with all the predictions 
of quantum theory, and with all the properties of relativistic quantum field 
theory. This will be shown later. Thus it provides for an enlargement of
the quantum-theoretical framework that allows one to consistently consider
alternative possible ``free choices.''

Any physicist is certainly free to deny this assumption that a later 
free choice cannot affect an already-observed-and-recorded outcome. 
Hence my proof must be regarded as an exploration of the logical 
consequences of making this foray into the realm of counterfactuals. 
This excursion is needed in order to formulate the concept of 
``no influence'' that is under consideration, and to give some
effective meaning to the notion of a ``free choice.''

Notice that there is no direct assumption that some unperformed
experiment has an outcome. In the first place, the effective
replacement assumption is that there is no influence backward 
in time. This assertion is part of the very assumption that is under 
examination, rather than a direct negation of a basic quantum precept. 
In the second place, the measurement whose outcome is counterfactually
specified to be $E+$ is exactly the locally defined experiment $E$
that is assumed to be the experiment that is actually performed at the 
earlier time. This earlier experiment and its outcome is ``unperformed'' 
or ``counterfactual'' only in a theoretical nonlocal and atemporal sense, 
when it is considered in combination with a never-to-be-performed later 
experiment. The no-backward-in-time-influence assumption is essentially 
the assumption that this earlier locally characterized measurement $E$
is one and the same measurement no matter which free choice is made 
later on. In this local sense there is no assumption about the outcome 
of any unperformed measurement: the single locally defined earlier 
measurement has one single outcome no matter which free choice is made 
later on.

Logicians have developed rigorous logical frameworks for consistent and 
unambiguous reasoning with statements involving counterfactual conditions. 
Those frameworks incorporate certain ideas about the world that are 
concordant with a deterministic classical-physics conception of nature. 
The present proof can be carried out within such a classical framework. 
However, the indeterminism of quantum physics leads naturally to a 
quantum logic for counterfactual reasoning. I shall, in Appendix A, present
a rigorous formal proof within this quantum counterfactual logic, which 
is described in Appendix B. 

This proof shows that  in a certain (Hardy) experimental 
setup a statement can be constructed whose truth value (true or false)
is {\it defined} in terms of the truth values of some statements
pertaining to possible outcomes of possible experiments confined to
a certain spacetime region $R$, but that this statement is, by virtue 
of certain predictions of quantum theory, {\it true} if one experiment 
is chosen and performed by an experimenter in an earlier (in LF) region 
$L$ that is spacelike separated from region $R$, but is {\it false} if 
a different experiment is chosen and performed in that earlier region $L$. 
This non-trivial dependence upon a free choice made in one region of the 
truth of a statement specified by the truth or falsity of statements 
pertaining to possible events in a spacelike separated region, constitutes,
within this logical framework, some kind kind of faster-than-light 
influence, as will be discussed.  

The quantum counterfactual logic described in appendix B is, 
I believe, interesting in itself, as is the formal proof given in
Appendix A. They give precision and rigor to the argument. However, 
the argument is so intuitively obvious that I believe it is sufficient
to state it in plain words. This is done in the next section. Then in 
section 3 it is shown, by referring to the Tomonaga-Schwinger formulation 
of quantum field theory, that the assumptions and conclusions of 
the proof are logically compatible with the predictions of relativistic 
quantum field theory. In section 4 I discuss the fact that the proved result 
is incompatible with the notion that the free choice made in one region 
can have {\it no influence of any kind} in a second region that is 
spacelike separated from the first. 

\vspace{.2in}
\noindent{\bf 2.  The Informal Proof.}
\vspace{.2in}

The argument is based on a Hardy-type [10] experimental set-up. 

There are two experimental spacetime regions $R$ and $L$, which are 
spacelike separated. In region $R$ there are two alternative possible 
measurements, R1 and R2. In region L there are two alternative possible 
measurements, L1 and L2. Each local experiment has two alternative possible 
outcomes, labelled by $+$ and $-$ . The symbol $R1$ appearing in a logical 
statement stands for the statement ``Experiment R1 is chosen and performed 
in region R.'' The symbol $R1+$ stands for the statement that ``The outcome 
`+' of experiment R1 appears in region R.'' Analogous statements with other
variables have the analogous meanings.
 
The detectors are assumed to be 100\% efficient, so that for each possible
world some outcome, either $+$ or $-$, will, according to quantum mechanics, 
appear in $R$, and some outcome, either $+$ or $-$,  will appear in $L$.

Suppose Robert acts in region $R$, and Lois acts in region $L$.
Then the first two predictions of QT for this Hardy setup are these:\\ 

(2.1): If Robert perform $R2$ and gets outcome $R2+$ and Lois performs $L2$,
       then Lois gets outcome $L2+$ .

(2.2): If Lois performs $L2$ and gets outcome $L2+$, 
       and Robert performs  $R1$ , then Robert gets 
       outcome $R1-$.\\

Combining these two conditions with the no-backward-in-time-influence
condition, which says that what Robert freely chooses to do in the
later region R cannot disturb Lois's earlier outcome, one immediately 
obtains the conclusion that:

If Lois performs $L2$ then\\
``If Robert performs $R2$ and gets $R2+$ then if his choice had gone
the other way he would have gotten outcome $R1-$.''

This conclusion is expressed by Line 5 of the formal proof.

The second two predictions of QT for this Hardy setup are:\\

(2.3): If Lois performs $L1$ and get outcome $L1-$, and Robert performs 
       $R2$, then Robert gets outcome $R2+$.
 
(2.4): {\it It is not true that} If Lois perform $L1$ and gets outcome 
       $L1-$, and Robert performs $R1$ then Robert gets outcome $R1-$.\\ 

To deduce the desired conclusion one uses the fact that if Lois performs
$L1$ then quantum theory predicts that she gets $L1-$ roughly half the time.
Thus there are physically possible worlds in which Lois performs
$L1$ and gets outcome $L1-$. In any such world if Robert chooses $R2$,
then, according to (2.3), he will obtain outcome $R2+$. According to
our no-backward-in-time-influence condition, the outcome $L1-$ observed
earlier by Lois would be left unchanged if Robert had, later in $R$,
made the other chioce, and performed $R1$.  But then prediction (2.4)
of quantum theory ensures that there are possible worlds in which
Lois performs $L1$ and Robert performs $R2$ and gets outcome $R2+$, but
in which Robert would not have obtained outcome $R1-$ of he had freely
chosen to perform $R1$ instead of $R2$. This means that 

If Lois performs $L1$ then {\it It is not true that}\\
``If Robert performs $R2$ and gets $R2+$ then if his choice
had gone the other way he would have gotten outcome $R1-$.''

This is exactly what the rigorous formal proof shows. 

Notice that this second conclusion, like the earlier one, contains 
the statement $SR$:\\

``If Robert performs $R2$ and gets $R2+$ then if his choice
had gone the other way he would have gotten outcome R1-.''

This statement $SR$, whose truth or falsity is defined in terms of the 
truth or falsity of statements pertaining to possible events in 
region R, is true if Lois's free choice in region $L$ is to perform $L2$, 
but is not true if Lois's free choice in region $L$ is to perform $L1$. 
Thus the truth of this statement $SR$ pertaining to region $R$ depends
upon what Lois freely chooses to do in a region $L$ that is situated 
spacelike relative to region $R$. This dependence constitutes
some kind of effect in region $R$ of Lois's free choice made in region
$L$. This is discussed in Section 4. 
\newpage
\vspace{.2in} 
\noindent{\bf 3. Logical Consistency and Compatibility with Relativity.}
\vspace{.2in}

The assumptions in this argument, unlike those of approaches based on 
hidden variables, are in line with the precepts of orthodox quantum theory, 
and when combined with the predictions of relativistic quantum field theory
they lead to no logical inconsistencies. One can confirm this by simply noting 
that the no-backward-in-time-influence condition is satisfied in the 
formulation of relativistic quantum field theory given by Tomonaga[11] and 
by Schwinger[12], with their spacelike surfaces $\sigma$ taken to be the 
constant-time surfaces in the special frame LF. This frame then defines the 
meaning of the evolving state $\Psi (t)$, which can be assumed to collapse 
to a new state when a measurement is completed, and new information is 
thus considered to have become specified. 

There is, of course, no suggestion in the works of Tomonaga and Schwinger 
that some particular set of surfaces should be singled out as the ``true'' 
or ``real'' surfaces that define the real evolving state of the universe
$\Psi (t)$ that is suddenly reduced to a new form when new information 
becomes available. Quite the opposite: they show that it does not matter 
which of the infinite collection of advancing sets of spacelike surfaces 
$\sigma $ one uses to define the forward-evolving state of the system. 
They effectively show that the predictions of the theory will be 
independent of which such set of advancing spacelike surfaces $\sigma$
one uses. This feature of their theory is, of course, completely 
concordant with the ideas of the theory of relativity. 

My use of Tomonaga-Schwinger theory is a logical, not ontological, one. 
I merely claim that my no-backward-in-time-influence assumption 
is {\it logically compatible} with the predictions of relativistic 
quantum field theory: I make no claim that this causality assumption 
has any ontological significance, or that the particular frame LF is 
unique. On the other hand, I could not demonstrate compatibility with 
the predictions of relativistic quantum field theory if I tried to 
assert that the no-backward-in-time-influence condition held 
simultaneously in several frames: Tomonaga-Schwinger theory does not 
ensure the compatibility of that stronger condition with the predictions 
of relativistic quantum field theory.

\vspace{.2in} 
\noindent{\bf 4. Conclusion.}
\vspace{.2in}

The two conclusions proved from our premises both involve the same 
assertion $SR$:

``If Robert performs $RA$ and gets $RA+$ then if his free choice in R
had gone the other way he would have gotten outcome $RC-$.''

Here $A$ stands for actual, and takes the value $2$, and $C$
stands for counterfactual, and takes the value $1$.

What was proved is that this statement $SR$ about the connections 
in $R$ of the consequences of making alternative possible free choices 
in $R$ is, by virtue of our explicity stated assumptions, true or false 
according to whether $L2$ or $L1$ is freely chosen in $L$. 

This conclusion entails that information about which choice 
is freely made in region $L$ must get to region $R$.
This is because the truth value (true or false) of $SR$ is defined 
in terms of the truth values of the elements of the quadruple of 
statement $(RA,RA+;RC,RC-)$: $SR$, with $A=2$ and $C=1$, is false if 
and only the set of truth values of that quadruple is $(t,t;t,f)$. For 
every other quadruple of truth values the statement $SR$ is true. 
Thus the nontrivial dependence of the truth of $SR$ on the free 
choice between $L1$ and $L2$ made in region $L$ means that the 
truth or falsity of some statements about possible events in region 
$R$ must, as a consequence of our assumptions, depend upon whether 
the free choice made in region $L$ is to perform $L1$ or $L2$. 

This result places a strong condition on theoretical models that reproduce 
the predictions of quantum theory. This condition is similar to the 
failure of locality associated with Bell's theorem. But here it is derived 
from the premises of ``free choice''  and ``no backward in time influence'' 
that are in line with the precepts of quantum theory, and that lead to no
logical contradictions, and to no conflicts with the predictions of 
relativistic quantum field theory.
\newpage 
\begin{center}
\vspace{.1in}
\noindent {\bf References}
\end{center}

1. Physics Today, December 1998, p.9.

2. W. Tittle, J. Brendel, H. Zbinden, and N. Gisin, 
        Phys. Rev. Lett. {\bf 81}, 3563 (1998).
 
3. P.G. Kwiat, E. Waks, A. White, I. Appelbaum, and Philippe Eberhardt,
      quant-ph/9810003. 

4. G Weihs, T. Jennewein, C. Simon, H. Weinfurter, and A. Zeilinger,
       Phys. Rev, Lett. {\bf 81}, 5039 (1998).

5. W. Tittle, J. Brendel, H. Zbinden, and N. Gisin, 
       quant-ph/9809025.

6. J.S. Bell, Physics, {\bf 1}, 195 (1964);  
    J. Clauser and A. Shimony, Rep. Prog. Phys. {\bf 41}, 1881 (1978).

7. J.S. Bell, {\it Speakable and Unspeakable in Quantum Mechanics},
    Cambridge Univ. Press, 1987, Ch. 4. 

8. H. Stapp, Epistemological Letters, June 1978. (Assoc. F Gonseth,
          Case Postal 1081, Bienne Switzerland).

9. A Fine, Phys. Rev. Lett. {\bf 48}, 291 (1982).

10. L. Hardy, Phys. Rev. Lett. {\bf 71}, 1665 (1993); P. Kwiat, P.Eberhard
A. Steinberg, and R. Chiao, Phys. Rev. {\bf A49}, 3209 (1994).

11. S. Tomonaga, Prog. Theor. Phys. {\bf 1}, 27 (1946).

12. J. Schwinger,  Phys. Rev. {\bf 82},  914 (1951).

13. H.Stapp, Amer. J. Phys. {\bf 65}, 300 (1997).

14. A. Shimony and H. Stein, in {\it Space-Time, Quantum Entanglement, and
Critical Entanglement: Essays in Honor of John Stachel.} eds. A. Ashtekar, 
R.S Cohen, D. Howard. J. Renn, S. Sakar, and A. Shimony  
(Dordrecht: Kluwer Academic Publishers) 2000.

\vspace{0.3in}

\newpage

\noindent {\bf APPENDIX A: The Formal Proof.}

Each line of the following proof is a strict consequence of the
predictions of quantum mechanics, (2.1)-(2.4), the general property
that there are possible worlds $W$ in which $L1$ is performed the outcome 
is $L1-$, the assumed property LOC1, and the properties of 
the rudimentary logical symbols. Line 6 is the one exception: it is just 
the same as line 5, but with L2 replaced by L1. The part of the proof 
from line 7 to line 14 shows that the statement on line 6 is false. 
Thus the proof shows that mentioned premises lead to the conclusion that 
line 5 is true, but that line 6 is false.

\noindent {\bf Proof}:

1. $(L2\wedge R2\wedge L2+)
   \Rightarrow [R1\Box\rightarrow (L2\wedge R1\wedge L2+)]$ \hfill [(B.6)]

2. $(L2\wedge R2\wedge R2+)\Rightarrow (L2\wedge R2\wedge L2+)$ \hfill [(2.1)]

3. $(L2\wedge R1\wedge L2+)
   \Rightarrow (L2\wedge R1\wedge R1-)]$ \hfill [(2.2)]

4. $(L2\wedge R2\wedge R2+)
   \Rightarrow [R1\Box\rightarrow (L2\wedge R1\wedge R1-)]$ 
    \hfill [1, 2, 3, (B.7)]

5. $L2\Rightarrow [(R2\wedge R2+)\rightarrow (R1\Box
   \rightarrow R1\wedge R1-)]$ \hfill [4, LOC1, (A.5)] 

6. $L1\Rightarrow [(R2\wedge R2+)\rightarrow (R1\Box
   \rightarrow R1\wedge R1-)]$ 

7. $(L1\wedge R2 \wedge R2+) \Rightarrow (R1\Box
   \rightarrow R1\wedge R1-)]$ \hfill [6, (A.5)]

8. $(L1\wedge R2\wedge L1-)\Rightarrow (L1\wedge R2\wedge R2+)$ 
    \hfill [(2.3)]

9. $(L1\wedge R2\wedge L1-)\Rightarrow  
   (R1 \Box\rightarrow R1\wedge R1-)$ \hfill [7, 8, (B.5)]

10. $(L1\wedge R2)\Rightarrow [L1- \rightarrow  
   (R1 \Box\rightarrow R1\wedge R1-)]$ \hfill [9, (A.5)]

11. $(L1\wedge R2)\Rightarrow 
   [R1 \Box\rightarrow (L1-\rightarrow R1\wedge R1-)]$ 
   \hfill [10, LOC1]
 
12. $(L1\wedge R1)\Rightarrow 
   \neg (L1-\rightarrow R1\wedge R1-)$ \hfill [(3.4)]   
   
13. $L1\Rightarrow [R1 \rightarrow \neg (L1-\rightarrow R1\wedge R1-)]$ 
    \hfill [12, (A.5)]

14. $(L1\wedge R2)\Rightarrow [R1\Box \rightarrow 
    \neg (L1- \rightarrow R1\wedge R1-)]$  \hfill [13, DEF.]
 
But the conjunction of $11$ and $14$ contradicts the assumption that the 
experimenters in regions  $R$ and  $L$ are free to choose which 
experiments they will perform, and that outcome $L1-$ sometimes occurs 
under  the conditions that  L1 and R1 are performed. Quantum theory
predicts that if L1 and R1 are  performed then outcome $L1-$
occurs half the time. Thus the falseness of the statement in line 6 is 
proved.  

[Note that there is only one strict conditional [$\Rightarrow $] in each
line. In an earlier brief description[13] of a theorem similar to the 
one proved above, but based on orthodox modal logic rather than the quantum
logic developed above, some material conditionals standing to the 
right of this strict conditional were mistakenly represented by the 
double arrow $\Rightarrow $, rather than by $\rightarrow$. 
I thank Abner Shimomy and Howard Stein [14] 
for alerting me to this notational error.]

The logical structure can be expressed in terms of {\it sets}.
This provides a compact method,  accessible to interested
physicists,  for validating the various lines of the proof.

For any statement $S$ expressed in terms of the rudimentary logical connections
let $\{W:S\}$ be the set of all (physically possible) worlds $W$ such that
statement $S$ is true at $W$ (i.e., $S$ is true in world W). Sometimes 
$\{W:S\}$ will be shortened to $\{S\}$.

A main set-theoretic definition is this: Suppose $A$ and $B$ are two
statements expressed in terms of the rudimentary logical connections.
Then $A\Rightarrow B$ is true if and only if 
the intersection of $\{A\}$ and $\{\neg B\}$ is void:
$$
[A\Rightarrow B] \equiv  [\{A\}\cap\{\neg B\}=\emptyset].  \eqno (A.1)
$$
Equivalently, $\{A\}$ is a subset of $\{B\}$:
$$
[A\Rightarrow B] \equiv  [\{A\} \subset \{B\}].  \eqno (A.2)
$$

Let $(S)_W$ mean that the statement $S$ is true at $W$. Then
$$
(A\rightarrow B)_W \equiv [(\neg A)_W \mbox{ or } (B)_W].     \eqno (A.3) 
$$
This entails that
$$
\{A\rightarrow B\}\equiv (\{\neg A\}\cup \{B\}). \eqno (A.4)
$$

\begin{center}
 {\bf Proof of (A.5)}
\end{center}

Equation (A.5) reads:
$$
[A\Rightarrow (B\rightarrow C)]\equiv [(A\cap B)\Rightarrow C]. \eqno (A.5)
$$
This is equivalent to
$$
[\{A\}\cap\{\neg (B\rightarrow C)\}=\emptyset]\\
\equiv [\{A\cap B\}\cap \{\neg C\}=\emptyset].    \eqno (A.6)
$$
But $\{\neg (B\rightarrow C)\}$ is the complement of 
$\{B\rightarrow C\}$. Using (A.4), and the fact that the complement
of $\{\neg B\} \cup \{C\}$ is $\{B\} \cap \{\neg C\}$, one obtains the
needed result.

\begin{center}
 {\bf Proof of line 5}
\end{center}

Line 4 has the condition L2 appearing to the right of the 
counterfactual condition R1. The  counterfactual condition 
R1 changes R2 to R1,  but leaves L2 unchanged. Hence the 
L2 appearing on the right can be omitted, since it appears 
already on the left. But then application of
(2.1) gives the line 5.

\begin{center}
 {\bf Proof of line 12}
\end{center}
Statement (3.4), expressed in  the set-theoretic form, says there is 
some world in $\{ L1\wedge R1\}\cap \{L1-\}$ that is not in $\{ R2-\}$. 
This entails that, in $\{ L1\wedge R1\}$, there some  world in 
$\{ L1-\}$ that is not in $\{R2-\}$. This is the form of (2.4) 
given in line 12.

\begin{center}
 {\bf Proof of line 14}
\end{center}

By definition, the assertion that $[C\Box \rightarrow D]$ is true in 
world $W$ is equivalent to the assertion that $D$ is true in {\it every} 
possible world $W'$ that differs from $W$ only by possible
effects of imposing condition $C$ rather than whatever condition in world 
$W$ is directly contradicted by condition $C$.

In line 14  the world $W$ can be any world in which L1 and R2 hold.
And $W'$ can be any world that differs from $W$ only by possible 
effects of changing R2 to R1. But no matter what these possible 
changes are, the world $W'$ must be a world in which L1 and R1 hold, 
and in any such world the statement $(L1-\rightarrow R1\wedge R1-)$ 
is false, by virtue of line 13. Thus line 14 is true.

\newpage
\vspace{.2in}
{\bf Appendix B. Quantum Logic for Counterfactuals}
\vspace{.2in}

Within orthodox quantum theory, with its notion of free choices on the part 
of the experimenters, and the notion of no-backward-in-time influence
of these free choices, logical reasoning covers unambiguously
some statements involving counterfactual conditions. The primary 
logical concept here is the notion of a ``logically possible world.'' 
It will be enough to define it in the case under consideration.

This case involves two spacelike-separated spacetime  regions  R and  L,
and in each region two alternative mutually exclusive experiments,
$R1$ and $R2$, and $L1$ and $L2$, respectively. Each possible experiment
has two possible outcomes, $R1+$ and $R1-$, etc. Thus there are
in this situation sixteen ``logically possible worlds'', each one 
labelled by one of the four globally defined experiments, 
$(R1,L1)$, $(R1,L2)$, $(R2,L1)$, and $(R2,L2)$, and by one of the 
four possible outcomes of that globally defined experiment, 
$(+,+)$, $(+,-)$,$(-,+)$, and $(-,-)$. This set of sixteen
logically possible worlds is the logical universe under consideration
here. Each such ``world'' might better be called a ``world history''.  
 
A {\it physically possible world} is logically possible world that
by virtue of the laws of nature (i.e., the predictions of quantum 
theory) has a non-null probability to occur. The physically possible 
worlds are called ``possible worlds.'' Normally, I omit also the 
word ``possible'': unless otherwise stated a ``world'' will mean 
a ``physically possible world''.

The rudimentary logical relationships involve the terms ``and'', 
``or'', ``equal'' and ``negation''. A rudimentary statement S involving 
these relations is said to be true in world W if and only if S is true 
by virtue of the set of conditions that define W and the laws of nature.

The concept of ``implication'' occurs, but it is important to 
distinguish between two different concepts.

The rudimentary relationship of implication is the so-called 
``material conditional''. It is defined in terms of the rudimentary 
logical relationships defined above, and it will be represented here 
by the single arrow $\rightarrow$. By definition:

 ``$ (A \rightarrow B) $ is true in world W'' is equivalent to

`` (A is false in W)  or (B is true in W) '' . \hfill (B.1) 

This rudimentary relationship is different from the logical relationship 
called the ``strict conditional'', which is represented here by the word 
``implies'', and a double arrow. The statement
 `` `A is true' implies `B is true' ''
is sometimes shortened to ``A implies B'', and is represented symbolically 
here by
$$
 A \Rightarrow B. \eqno (B.2)
$$

By definition, $A\Rightarrow B$ is true if and 
only if for {\it every} (physically possible) world W either  
``A is false in W'' or   ``B is true in W'': i.e.,  for {\it every} 
(physically possible) world W, the rudimentary statement 
$(A\rightarrow B)$ is true in W. 

The proof to be presented here is based on a causality condition called LOC1.
It expresses the condition that there is at least one Lorentz 
frame, LF, such that if an experiment is performed and the outcome is 
recorded $\it prior$ to some time $T$, as measured in LF, then this 
outcome can be regarded as fixed and settled, independently of which 
experiment will eventually be freely chosen and performed (faraway) 
at a time later than $T$. 

It is assumed that the regions L and R lie earlier and later that this time
$T$, respectively. Then LOC1 means that if Lois (acting in L)  performs her 
experiment before Robert (acting in R), we can safely assume that her result 
does not depend on what Robert will do, but not vice-versa.

Logicians deal with statements of this kind by employing a 
third kind of implication. It uses the concept ``instead of''.  
This concept is of central importance in classical counterfactual reasoning, 
and it has an unambiguous meaning within our quantum context. 
That meaning is now explained.

Suppose that $A$ represents some possible conditions that the experimenters 
could  set up, and some conditions on the possible outcomes. [For example,
$A$ could be the condition that Lois and Robert perform $L2$ and $R2$,
respectively, and that Lois gets outcome $L2+$]

Suppose condition $C$ represents some free choice (by some experimenter) that 
could conflict with $A$. [In the example, $C$ could be 
``Robert performs $R1$'']. 

Finally, suppose condition $D$ represents some possible outcome that could 
occur if $C$ were to hold ``instead of'', whatever condition $C$ contradicts. 
[In the example, $D$ could be ``Lois gets outcome $L2+$,'' or perhaps 
``Robert gets outcome $R1-$.'']

Then consider a statement of the form:\\
  \\
 ``$A$ implies [If, instead, $C$ then $D$]'' \hfill (B.3)\\
  \\

The phrase ``If, instead, $C$ then $D$'' is traditionally represented 
symbolically by $[C \Box \rightarrow D]$, and I shall use that symbolic form 
for the quantum version defined here. Like all rudimentary statements 
the assertion $[C\Box \rightarrow D]$ it is a statement that is made 
{\it in} one world, say $W$. But it is a statement {\it about} about an 
entire set of worlds $W'$, namely the set of all (physically possible) worlds 
that differ from $W$ only by possible consequences of choosing the 
experimental condition $C$  instead of whatever condition in world  $W$ 
conflicts with $C$.

Given this  definition of  $[C\Box \rightarrow D]$  the statement
$$
 A \Rightarrow [C \Box \rightarrow D] \eqno(B.4)
$$
expresses the condition that, for all (physically possible) worlds $W$, if 
$A$ is true in $W$ then then $D$ is true in every (physically possible) 
world $W'$ that differs from $W$ only by possible effects of choosing 
condition $C$ instead of whatever condition in $W$ conflicts with $C$.

It is essential that this definition allows this statement to be combined
with other logical statements in an unambiguous way. In particular, the 
usual laws of logic can be applied, {\it without any change}, to arguments 
involving statements of this kind. Suppose, or example, that one has, in 
addition to the truth of (B.4), also the truth of $(B\Rightarrow A)$, which 
asserts that, for all $W$, if $B$ is true in $W$ then $A$ is true in  $W$.  
Then one can immediately conclude from  the  meaning  of $(B.4)$ that
$$
B \Rightarrow [C \Box \rightarrow D]. \eqno(B.5)
$$

The definition  of [C $\Box \rightarrow$ D] is general. But in order 
to make use of it one must have some  condition on the 
``possible effects of choosing condition $C$ instead of whatever 
condition in world $W$ conflicts with  $C$.''

This is where LOC1 comes in. Suppose the region $L$, where Lois acts,
lies earlier than the region $R$, where Robert acts. Then LOC1 entails 
[with ``and'' represented by $\wedge$]  

$$
(L2\wedge R2\wedge L2+)
   \Rightarrow [R1\Box\rightarrow (L2\wedge R1\wedge L2+)] \eqno(B.6)
$$

This statement is true by virtue of the LOC1 premise that the outcome
that Lois gets, and also her free choice, do not depend on what Robert 
does later.

Another example of the logical rules is this. Suppose that (B.5) 
is true. And suppose that $F$ is a condition on the outcome under 
the alternative condition $C$, and that $D \Rightarrow F$ is true. 
This is the condition that, for every $W'$, if $D$ is true in $W'$  
then $F$ is true in $W'$. Then the meaning of (B.5), as described above
(B.4), with B in place of A, ensures that the following statement 
is  true: 
$$
 B \Rightarrow [C \Box \rightarrow F ]. \eqno(B.7)
$$
This result is used to get line 4 of the proof given in Appendix A.

All the other lines of the proof given in Appendix A can be 
strictly deduced, in a similar way, from just the meanings of the logical
symbols, the predictions of quantum theory, and the property LOC1.

\end{document}